\documentclass{article}

\usepackage{arxiv}

\usepackage[utf8]{inputenc} 
\usepackage[T1]{fontenc}    
\usepackage{hyperref}       
\usepackage{url}            
\usepackage{booktabs}       
\usepackage{amsfonts}       
\usepackage{nicefrac}       
\usepackage{microtype}      
\usepackage{cleveref}       
\usepackage{lipsum}         
\usepackage{graphicx}
\usepackage{natbib}
\usepackage{doi}

\title{Gray Swan Factory: Making Extreme Events from Ordinary Cyclones}

\date{31 March 2026}

\newif\ifuniqueAffiliation

\ifuniqueAffiliation 
\author{ \href{https://orcid.org/0000-0000-0000-0000}{\includegraphics[scale=0.06]{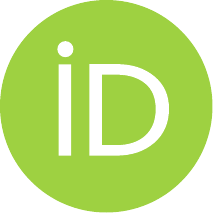}\hspace{1mm}David S.~Hippocampus}\thanks{Use footnote for providing further
		information about author (webpage, alternative
		address)---\emph{not} for acknowledging funding agencies.} \\
	Department of Computer Science\\
	Cranberry-Lemon University\\
	Pittsburgh, PA 15213 \\
	\texttt{hippo@cs.cranberry-lemon.edu} \\
	\And
	\href{https://orcid.org/0000-0000-0000-0000}{\includegraphics[scale=0.06]{orcid.pdf}\hspace{1mm}Elias D.~Striatum} \\
	Department of Electrical Engineering\\
	Mount-Sheikh University\\
	Santa Narimana, Levand \\
	\texttt{stariate@ee.mount-sheikh.edu} \\
}
\else
\usepackage{authblk}

\newbox{\orcid}\sbox{\orcid}{\includegraphics[scale=0.06]{orcid.pdf}} 
\author[1]{%
	\href{https://orcid.org/0000-0001-8486-9739}{\usebox{\orcid}\hspace{1mm}Gregory J. Hakim \thanks{\texttt{ghakim@uw.edu}}}%
}
\author[2]{%
	\href{https://orcid.org/0009-0006-7242-2351}{\usebox{\orcid}\hspace{1mm}Aishwarya Agrawal}%
}
\affil[1]{Department of Atmospheric Sciences, University of Washington, Seattle, WA, USA}
\affil[2]{Interlake High School, Bellevue, WA, USA}
\fi

\renewcommand{\shorttitle}{Gray Swan Factory}

\hypersetup{
pdftitle={Gray Swan Factory: Making Extreme Events from Ordinary Cyclones },
pdfauthor={Gregory~J.~Hakim, Aishwarya~Agrawal},
pdfkeywords={Gray Swans, Extreme Events, Cyclones, Hurricanes},
}

\begin{document}
\maketitle

\begin{abstract}

Gray swans, plausible but unobserved extreme events, broaden our understanding of the range of hazards beyond those observed during the short observational record. They are useful for dynamical studies, synthetic training data, emergency planning, infrastructure design, and insurance hazard assessment. We propose a method to produce gray swans from the observational record using gradient descent on a loss function with a differentiable weather prediction model. Minimizing the loss corresponds to perturbed initial conditions that produce a measurable outcome at a future time, subject to constraints, such as the size of the initial perturbations. We illustrate the method by altering hurricane Fiona (2022), which tracked northward over the Atlantic Ocean, to produce a gray-swan outcome similar to hurricane Sandy (2012), which made landfall on the East Coast of the United States after a unique westward turn. The Fiona gray-swan solution, involving small perturbations to reanalysis initial conditions, produces an extratropical cyclone with a Sandy-like track, a warm core, and a minimum sea-level pressure more than 20 hPa lower than Sandy. Perturbations to the extratropical state are more important than to the hurricane, leading to interactive strengthening, and merger, of an upper-level trough and the hurricane. Similar gray swans are found for four other Atlantic hurricanes. A major weakness of this work is that the hurricane core is not resolved by the model used for optimization, and the impact of this is unknown. Furthermore, although these solutions present plausible outcomes, they do not inform on their probability of occurrence.
  
\end{abstract}


\section{Introduction \label{sec:intro}}

Sampling extreme weather events is critical for a wide range of scientific and societal problems, from understanding the dynamics of these event to pricing risks associated with them. Given the short observational record, extreme events are, by definition, rare. Black swans are extreme events that are unknown prior to their occurence \citep[e.g.,][]{aven2015implications,masys2012black}. Gray swans are potentially much more useful, because they are known, plausible outcomes, that have not yet occurred \citep{lin2016grey,horsburgh2021grey,sun2025can}.They may be useful for emergency planning, infrastructure design, reinsurance hazard assessment, and tests of dynamical sensitivity for weather extremes. Finding gray swans typically involves numerical simulations, possibly for very long periods of time, or for large ensembles at one time, in order to generate larger sample sizes than what is available from the observational record. This approach is computationally intensive, even with efficient machine learning models, since it is an unguided random sampling strategy that visits tail events infrequently. Here we propose to use gradient descent on an objective function to ``manufacture'' storms to the specifications encoded in the objective function. The example application is hurricane Sandy (2012), which was a devastating storm on the Mid Atlantic coastline of the United States \citep{halverson2013hurricane,galarneau2013intensification,shen2013genesis,kunz2013investigation,lackmann2015hurricane,shin2017impact,munsell2014prediction,mattingly2015climatological,kowaleski2018relationship,qian2016examination}. We explore whether ``ordinary'' storms can be guided to a Sandy-like outcome using gradient descent.

Our gradient descent approach is a nonlinear extension of adjoint sensitivity, and has recently been applied to computing shadowing trajectories to test the limit of atmospheric predictability \citep{vonich2024,vonich_hakim26}. Traditional adjoint sensitivity \citep[e.g.,][]{langland1995,errico1997adjoint,langland2002,doyle2012,doyle2014,doyle2019,lloveras2025_goose} typically uses a tangent-linear model (TLM), which linearizes the full nonlinear model to evolve the dynamics of small perturbations. The adjoint version of the TLM is integrated backward in time, to propagate sensitivity gradients of a functional on the end state (a metric) to the initial conditions. One can then evaluate the response of the forecast in the metric to specific changes in the initial conditions that most effectively change the forecast by perturbing in the direction of the gradient. The results depend on the choice of metric, and the range of perturbation amplitudes over which the linearization applies. One can make several forward--backward passes with the TLM--adjoint system to change the metric, but at some point a new nonlinear forecast must be made, and the linearization process repeated. With traditional models this ``inner and outer loop'' optimization is expensive and involves developing the TLM and adjoint models from the source code of the nonlinear model for explicit integration.

A new generation of models has emerged in recent years that has two novel aspects: (1) they are extremely computationally efficient; and (2) they are fully differentiable through automatic differentiation engines in modern machine-learning frameworks. Computational efficiency addresses the traditional inner--outer loop bottleneck by removing the need for the inner loop, since the nonlinear forward pass is much cheaper than the TLM of traditional models. Full differentiability means that the adjoints of these models are efficiently evaluated by algorithmic applications of the chain rule as compared to adjoints of traditional models, which involve explicit code to express the TLM and adjoint models. Together, these two novel aspects promote deep gradient-descent searches, by which we mean hundreds to thousands of iterations (``epochs'') combining the forward and backward passes using the fully nonlinear model and automatic differentiation. This search defines the reachability set \citep[e.g.,][]{althoff2021} of gray swan outcomes as a constrained optimization problem regularized to remain ``near'' an observed state, such as reanalysis data.

This algorithm has recently been applied by \citet{whittaker2026} as a storyline approach to the 2021 Pacific Northwest heatwave. In particular, they found optimal initial conditions that produce a heatwave 3.7K warmer that the most extreme outcome in a 75-member ensemble. \citet{whittaker2026} emphasize the efficiency of the optimization approach, as compared to large ensembles, for exploring rare events. We build on this idea here by applying it to the production of gray swans, rather than amplifying an existing extreme event.

Constructing a gray swan factory involves a specific application of this efficient gradient-descent algorithm to achieve the objectives of the manufacturer. We express this objective as a loss function, described in section \ref{sec:method} with the goal of creating outcomes like hurricane Sandy (2012) from ordinary, observed, hurricanes. The gray-swan factory is then illustrated for the Atlantic hurricane Fiona (2022) \cite{malinina2025attribution}, which followed a typical recurvature trajectory from the tropical Atlantic northward to the Canadian maritime provinces. In section \ref{sec:results} we show that the gray-swan search algorithm finds a nearby initial condition that achieves the objective by producing a storm like hurricane Sandy (2012). We test the sensitivity of the algorithm to a variety of design-choice parameters in section \ref{sec:sensitivity}. In section \ref{sec:other_storms}, we show the results of the gray-swan search algorithm for four additional storms drawn from the recent past. Conclusions are provided in section \ref{sec:conclusions}.

\section{Method}
\label{sec:method}

Gray swans are part of a broader class of initial-value reachability problems \citep[e.g.,][]{althoff2021} defined by perturbing a reference initial condition in order to achieve a forecast objective at a later time, $t$. For gray swans, the objective is to find plausible extreme events that have not previously occurred, and are derived as small perturbations to observed initial conditions. Here we take the reference initial conditions from the ERA5 reanalysis \citep{hersbach2020era5}, and the forecast objective is a storm at a later time similar to hurricane Sandy (2012). The simplist scalar loss that defines this objective is a measure of storm intensity at a single point having (longitude, latitude) = $(x_0,y_0)$. Here we choose the 1000hPa geopotential height at the model grid point closest to hurricane Sandy's landfall.  Since the range of potential outcomes scales with the length of the forecast time window, we only consider relatively short forecasts of 2--4 days. Similarly, the range of potential outcomes scales with the magnitude of the initial condition perturbations, so we include a control parameter that promotes smaller perturbations:
\begin{equation}
J\, =\,  J_f + J_i + J_t = (x_G - g)^2 + \frac{\alpha_i}{N}\sum ({\mathbf x}_T - \overline{\mathbf x}_T)^2  + \frac{\alpha_t}{N}\sum ({\mathbf x}^{+}_G - {\mathbf x}_G)^2. 
  \label{eqn:loss}
\end{equation}
\noindent Symbols represent scalars unless in bold. $x_G$ is the forecast 1000 hPa geopotential height at $(x_0,y_0, t)$ and $g$ is a target value. Sums for $J_i$ and $J_t$ are taken over a set of $N$ points in the model state where temperature and geopotential are defined, denoted by subscripts $T$ and $G$, respectively. An overbar represents the reference initial condition (ERA5). Contribution $J_t$ penalizes the mean-squared tendency magnitude of the first model forecast time step (one hour), ${\mathbf x}^{+}_G$, measured in the geopotential field. Without this penalty, the minimization process does not control rapidly decaying noise in the initial field, which we find accumulates during the minimization process in the absence of this penalty. Loss (\ref{eqn:loss}) is similar to the one used by \citep{whittaker2026}, although they did not include $J_t$, and $J_f$ has a different objective in their study.

Temperature and geopotential height variables are standardized by dividing the full values by the spatial standard deviation of the ERA5 field at the initial time on each isobaric level. The ``standard experiment'' described below uses the values $g=-1000$m (before standardization), $x_0 \approx 73^{\circ}$W, $y_0 \approx 38^{\circ}$N, $\alpha_i = 10^4$, and $\alpha_t = 10^2$. Sensitivity to $\alpha_i$ will be discussed below, and we note that although $g$ is subjectively chosen, a value that is too large limits the gray swan intensity to values above that choice.

To minimize the loss function we perform numerical gradient descent on (\ref{eqn:loss}) using version-1 of the hybrid model NeuralGCM \citep{kochkov2024neural}(NGCM), which has a traditional spherical harmonics solver for vorticity, divergence, temperature anomaly, dry air mass, and three species of water (vapor, cloud water, and cloud ice). The state evolves by numerical integration, with occasional corrections from a learned model that accumulates the contributions from unrepresented and unresolved processes, and is trained by minimizing forecast errors measured on ERA5. The model is available in three spatial resolutions, each having a different learned component: 2.8$^{\circ}$, 1.4$^{\circ}$, and 0.7$^{\circ}$. Optimization is performed here using the 2.8$^{\circ}$ model, which has been modified from the original version to conserve global dry air mass. Moreover, we set the gradients of all water variables to zero during minimization, which allows for larger learning rates; these fields adjust rapidly during the forecast cycle in response to changes to the other fields.

Derivatives are evaluated using JAX, and Optax is used for managing the descent process \citep{deepmind2020jax}. We use the Adam optimizer with a learning rate that increases linearly from $10^{-6}$ to $10^{-5}$ over the first 10 epochs, and constant learning rate thereafter; all other parameters for the optimizer are left at default values. All optimizations are run for 1000 epochs on an NVIDIA GB10 GPU having 128GB of VRAM.

The first storm we consider is hurricane Fiona (2022), which formed over the subtropical Atlantic, and recurved northward over the western North Atlantic, making landfall as an extratropical system over eastern Nova Scotia \citep{malinina2025attribution}. The standard experiment performs the gray swan optimization starting at 00 UTC 22 September 2022 (22/00UTC; hereafter, dates and times are abbreviated as day/hour~UTC) targeting a Sandy (2012)-like outcome at 24/00UTC as measured by $J_f$. This time interval is subjectively chosen, and the affect of that choice will be shown below by optimizing the loss for different starting times. NGCM output variables are limited to isobaric surfaces, and when plotting near-surface fields we use the lowest level, 1000 hPa. Moreover, instead of plotting the geopotential height at 1000 hPa, we convert that field to equivalent mean-sea-level pressure (MSLP) by reduction to sea level using the US Standard Atmosphere. All fields are interpolated for plotting to 0.7$^{\circ}$ degree resolution by zero-padding spherical harmonics.

\begin{figure}[ht!]
\begin{center}
  \noindent\includegraphics[width=.65\textwidth]{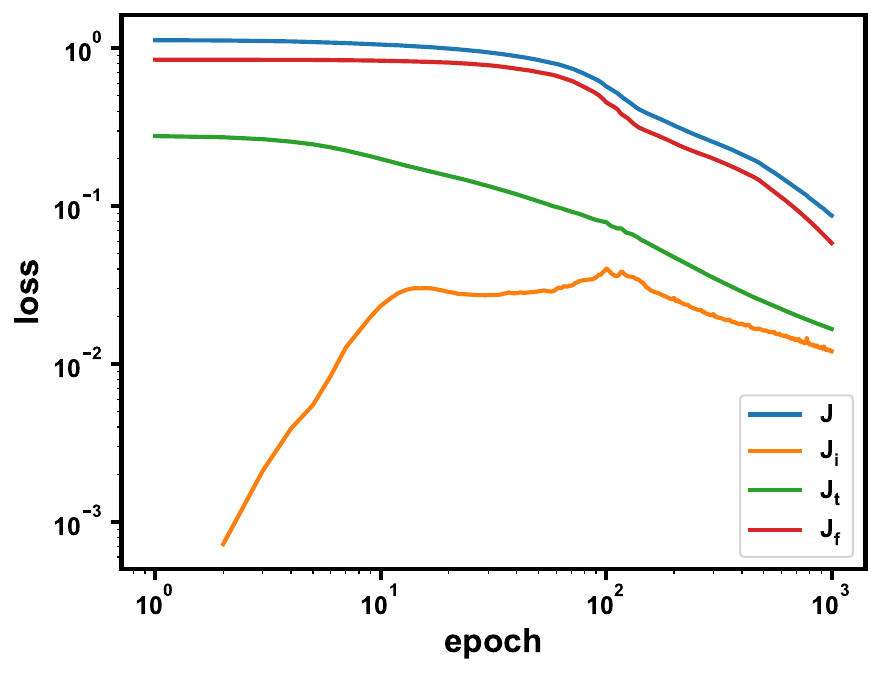}
\end{center}
  \caption{Loss function and contributions from each component as a function of optimization epoch: $J$, total loss (blue); $J_f$, final-time loss (red); $J_i$, initial-time loss (orange); $J_t$, tendency loss (green).} \label{fig:loss_breakdown}
\end{figure}

The loss evolves through stages, with the first involving a steep drop around $\sim$50--100 epochs, followed by a slower power law, which steepens in the last few hundred epochs  (Fig. \ref{fig:loss_breakdown}, blue line). As we show in section \ref{sec:results}, the first stage involves changes to the initial condition that relocate the forecast storm position from Nova Scotia to the Mid Atlantic coast, and thereafter changes that affect the intensity of the storm at the target location. The primary contribution to the loss for all epochs is $J_f$ (red line). The tendency penalty, $J_t$ is largest at the beginning and decreases at an increasing rate (green line). The penalty on the size of the initial condition, $J_i$ initially follows the ramping learning rate of the optimization, reaching a maximum around 100 epochs, and then gradually decreases (orange line).

\begin{figure}[ht!]
\begin{center}
  \noindent\includegraphics[width=.65\textwidth]{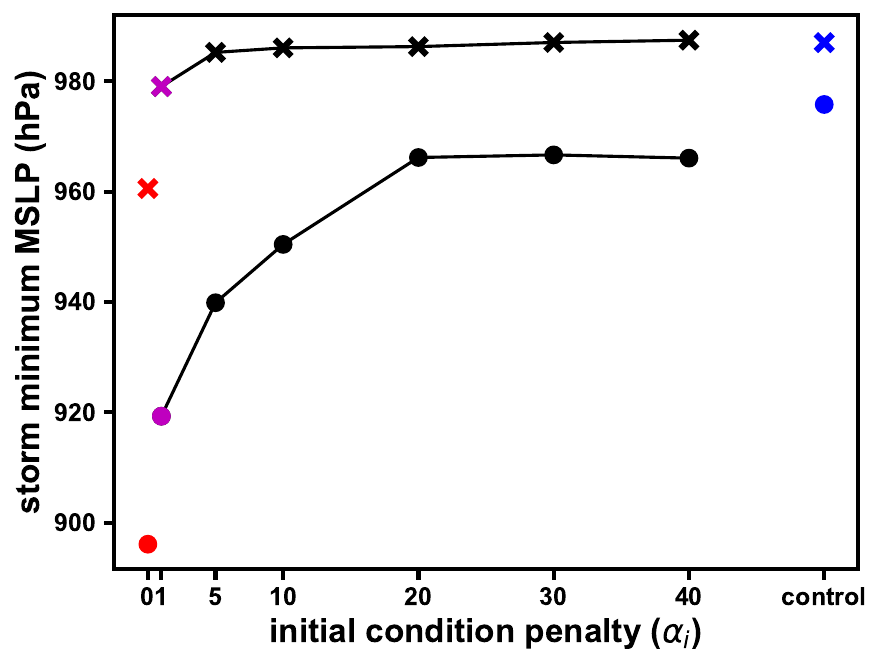}
\end{center}
  \caption{Minimum MSLP as a function of the loss-function initial-condition penalty, $\alpha_i$ for the gray swan initial condition (crosses) and solution at the optimization time (circles). Values for the standard experiment are shown in magenta, for the control in blue ($\alpha_i \rightarrow \infty$), and $\alpha_i=0$ in red. $\alpha_i$ values on the abscissa are scaled relative to the standard experiments (i.e., a value of unity is $10^4$) .} \label{fig:mslp_ic_penalty}
\end{figure}

An important sensitivity of the solution derives from the penalty on the size of the initial condition, $J_i$. As measured in the minimum MSLP at forecast time $t$ over the western Atlantic ocean, the optimization limits are bounded below by $\alpha_i \rightarrow 0$ (no penalty) and above by $\alpha_i \rightarrow \infty$, which has zero perturbations and thus returns the control forecast (Fig.\ref{fig:mslp_ic_penalty}). The standard experiment has $\alpha_i = 10^4$, which yields a MSLP of 919 hPa (magenta circle) at $(x_0,y_0)$, compared to  976 hPa for the control (blue circle) (with a storm located near Nova Scotia). Increasing $\alpha_i$ by a factor of 5 gives a minimum MSLP value of 940hPa, which is around the value of hurricane Sandy (2012), and a factor of 10 gives 950 hPa; sensitivity is weak for values greater than 20 times the standard experiment. Setting $\alpha_i=0$ allows for unlimited changes to the initial condition, and a minimum MSLP of 896 hPa. Sensitivity of the intensity of the initial hurricane in minimum MSLP to $\alpha_i$ is much smaller than for the solution 48~h later (Fig.\ref{fig:mslp_ic_penalty}, crosses), with little change above $\alpha_i \approx 5$. We emphasize that the choice of $\alpha_i$ is arbitrary and up to the user; it may also include spatially varying penalties, which we have not considered here.

\section{Gray swans for Fiona (2022)}
\label{sec:results}

\begin{figure}[ht!]
\begin{center}
  \noindent\includegraphics[width=1.0\textwidth]{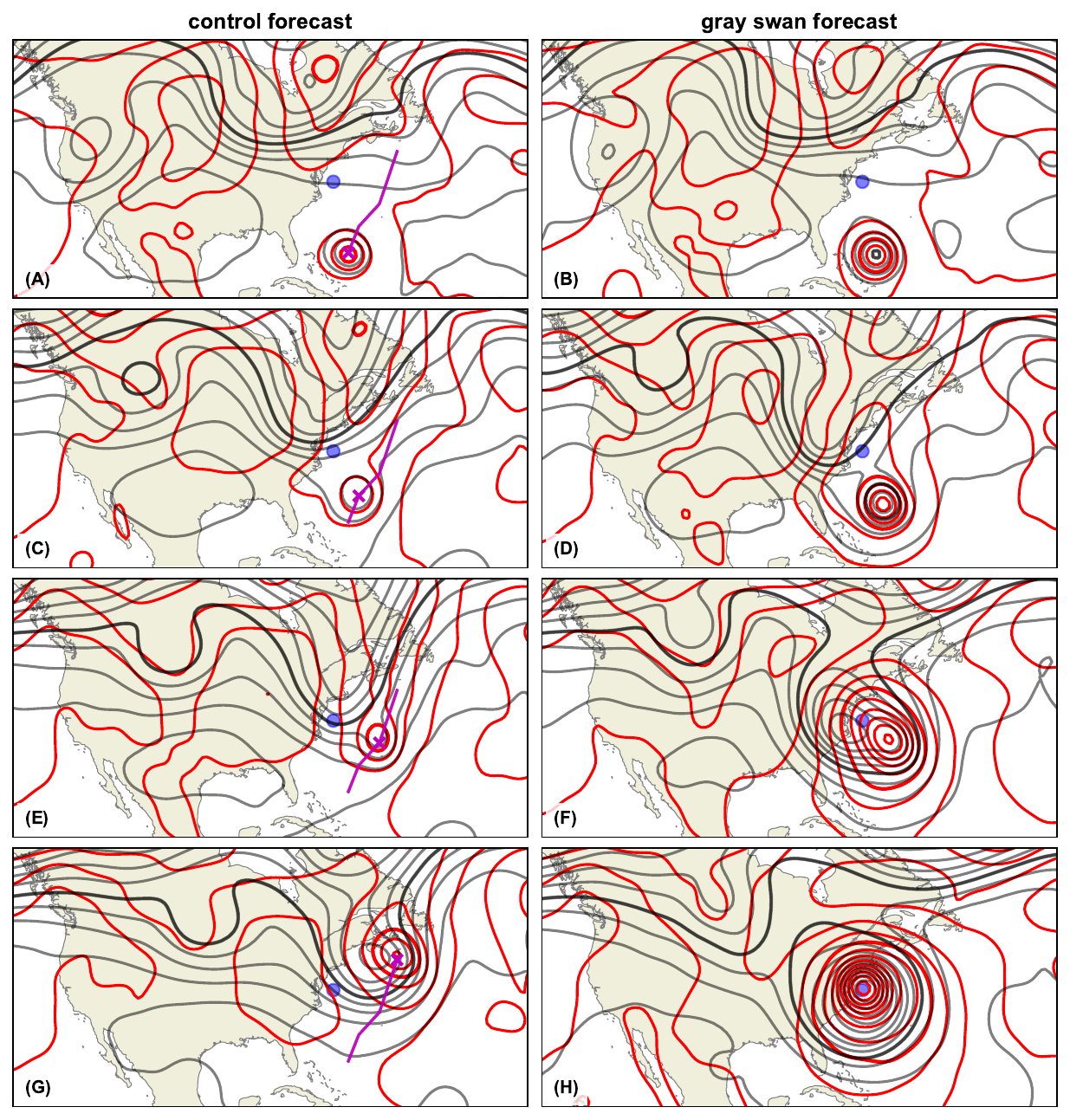}
\end{center}
  \caption{Control (left column) comparison with gray swan optimal solution (right panel) as a function of time: (A, B) 2022-09-22T06, (C, D) 2022-09-23T00, (E,F) 2022-09-23T12, (G, H) 2022-09-24T00. MSLP is shown in red contours every 8 hPa, with the 992 hPa contour in bold; 500 hPa geopotential height is shown in gray contours every 60m with the 5640m contour in bold. NOAA NHC best-track location is shown in the magenta line, with ``X'' symbols marking the location of the storm at the time of each panel. The blue dot shows the optimization location at 00 UTC 24 September 2022.} \label{fig:control_optimal_lead_time}
\end{figure}

The evolution of the MSLP and 500 hPa geopotential height field for the gray swan solution for hurricane Fiona (2022) using the standard experiment configuration is shown in Fig.~(\ref{fig:control_optimal_lead_time}). For the control forecast, initialized with ERA5, the storm closely follows the observed track (magenta line and `X' symbols on Fig.~\ref{fig:control_optimal_lead_time}, left panels). Forecast storm intensity is greatly underestimated, with simulated MSLP minima around 985--990 hPa until 12UTC 23 September when the pressure falls to 984 hPa, and then to 976 hPa 12h later. In the best-track data the storm intensity during the two-day time period shown in Fig.~\ref{fig:control_optimal_lead_time} is nearly constant around 930--935 hPa. At 2.8$^{\circ}$ horizontal resolution, the model is unable to resolve any details of the hurricane aside from it being a warm-core cyclone.

For the the gray swan forecast, the storm track is similar to the control up to 23/12UTC, after which the storm turns sharply westward toward the mid-Atlantic coastline, similar to hurricane Sandy (Fig.~\ref{fig:control_optimal_lead_time}, right panels). The gray-swan storm is also more intense than in the control (979 hPa at 22/06UTC compared to 986 hPa for the control), and deepens steadily with time, reaching 952 hPa at 23/12UTC. Very rapid deepening characterizes the final 12 h, with the minimum MSLP decreasing 33 hPa in 12 h to 919 hPa. This is considerably deeper than hurricane Sandy, which had a minimum pressure of 940 hPa. The essential difference between the gray swan and control solutions involves the interaction between the hurricane and the midlatitude trough in the 500hPa height field. This is evident at 23/00UTC, when the trough in the gray swan solution is shifted west, with a shorter zonal wavelength, compared to the control (Fig.~\ref{fig:control_optimal_lead_time}C,D). After this time the hurricane and trough mutually interact by co-rotating and merging into a single vertically aligned vortex (Fig.~\ref{fig:control_optimal_lead_time}H); in the control this interaction is weaker, and the two systems remain distinct (Fig.~\ref{fig:control_optimal_lead_time}G).

%

\begin{figure}[ht!]
\begin{center}
  \noindent\includegraphics[width=1.0\textwidth]{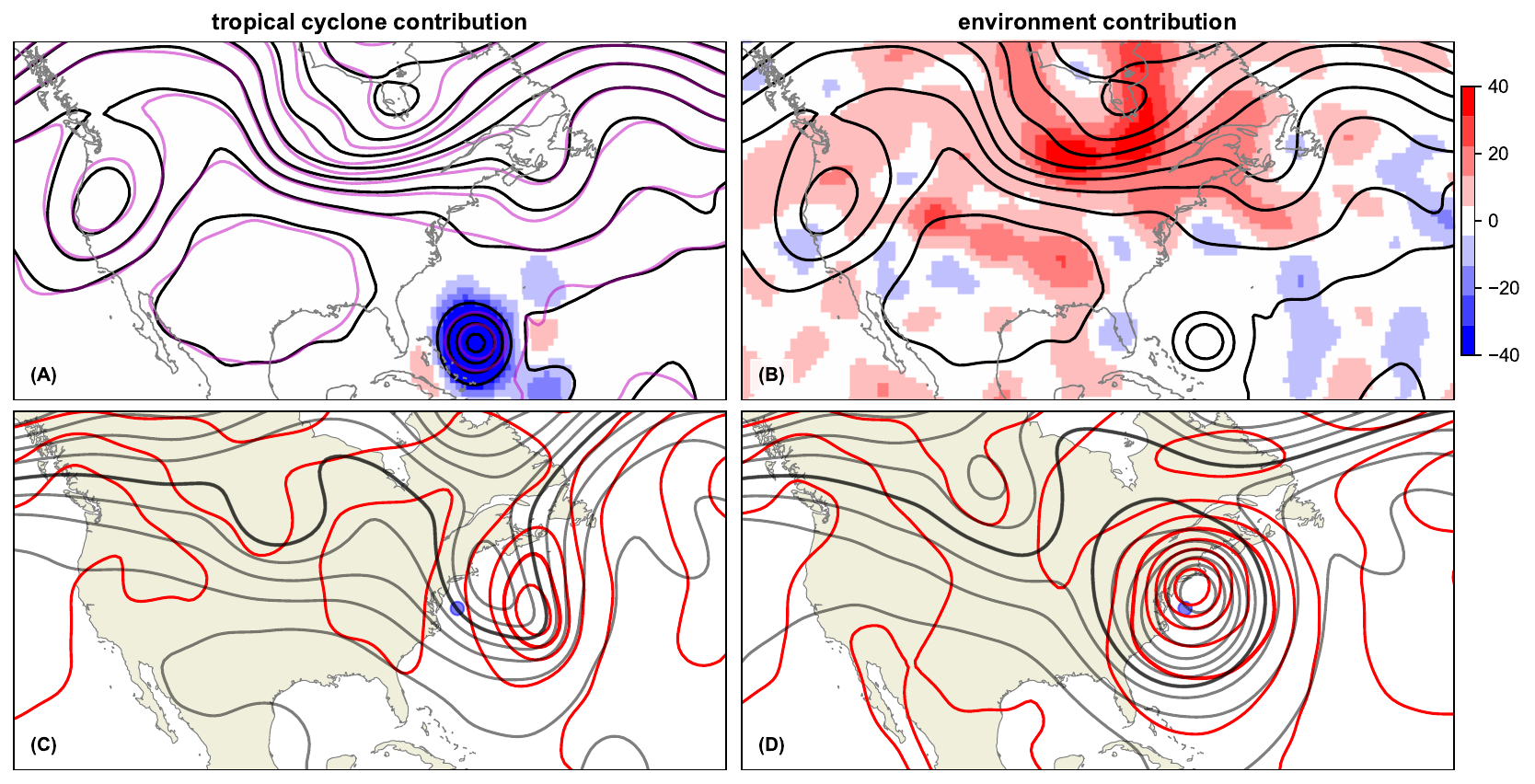}
\end{center}
  \caption{Initial conditions (22/00UTC; panels A,B) and forecasts (24/00UTC; panels C,D) for optimal initial perturbations limited to the tropical cyclone (A,C) and environment (B,D). Optimal initial 500hPa geopotential height is shown by black lines every 60m in (A,B), and the control by magenta lines in (A). Colorfill in (B) is the difference field (optimal-control) in 500hPa geopotential height (m). In (C,D), MSLP is shown in red contours every 8 hPa, with the 992 hPa contour in bold, and 500 hPa geopotential height is shown in gray contours every 60m with the 5640m contour in bold. The tropical cyclone domain is defined by the region bounded within 15$^{\circ}$--35$^{\circ}$N and 60$^{\circ}$--80$^{\circ}$W.} \label{fig:optimal_TC_EN}
\end{figure}

Further evidence for the importance of the extratropical contribution to the gray swan is provided in Fig.~\ref{fig:optimal_TC_EN}, which shows solutions for gray swan initial perturbations limited to the tropical cyclone (left panels), and everywhere else (``environment''; right panels). When initial perturbations are limited to the tropical cyclone, the solution is similar to the control solution, but with a delay in the northward movement of the storm (Fig.~\ref{fig:optimal_TC_EN}C). In contrast, the solution for initial perturbations in the environment is closer to the full gray swan, but weaker, with a minimum MSLP of 954 hPa. Comparing the control and gray swan 500 hPa height fields (magenta and gray contours in Fig.~\ref{fig:optimal_TC_EN}A, respectively), and the difference field (colorfill in Fig.~\ref{fig:optimal_TC_EN}A,B) reveals the largest extratropical changes near the upstream trough, raising the heights to weaken the trough and shift it slightly westward. Note that, because the solution is fully nonlinear, the sum of  Fig.~\ref{fig:optimal_TC_EN}C,D is not the same as the full optimal, Fig.~\ref{fig:control_optimal_lead_time}H, but is a rough approximation (not shown).

\begin{figure}[ht!]
\begin{center}
  \noindent\includegraphics[width=1.0\textwidth]{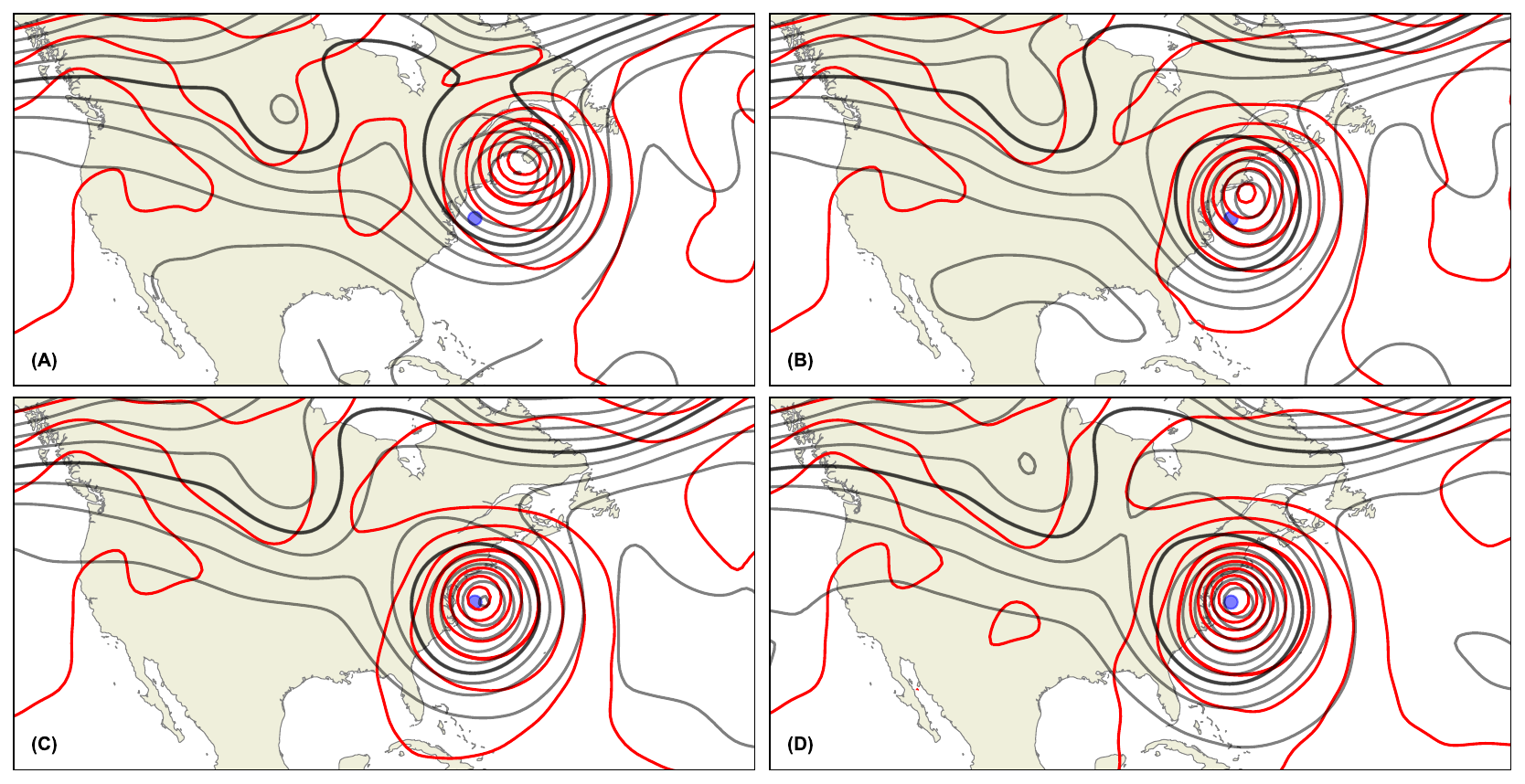}
\end{center}
  \caption{Gray swan solution at 24/00UTC as a function of optimization epoch: (A) 50, (B) 100, (C) 200,and (D) 400. MSLP is shown in red contours every 8 hPa, with the 992 hPa contour in bold, and 500 hPa geopotential height is shown in gray contours every 60m with the 5640m contour in bold.} \label{fig:optimal_epoch}
\end{figure}

Fig.~\ref{fig:optimal_epoch} reveals how the gray swan solution emerges smoothly from the control initial condition as a function of optimization epoch. In the first 100--200 epochs, the storm shifts westward, and then southwestward (Fig.~\ref{fig:optimal_epoch}A,B,C). This is the portion of the optimization when the steepest decrease in the loss is observed (cf. Fig.~\ref{fig:loss_breakdown}). Beyond 200 epochs, the storm remains in the target location, and initial-condition changes contribute to storm intensification at the target time (Figs.~\ref{fig:optimal_epoch}D, \ref{fig:control_optimal_lead_time}H ).

\begin{figure}[ht!]
\begin{center}
  \noindent\includegraphics[width=.65\textwidth]{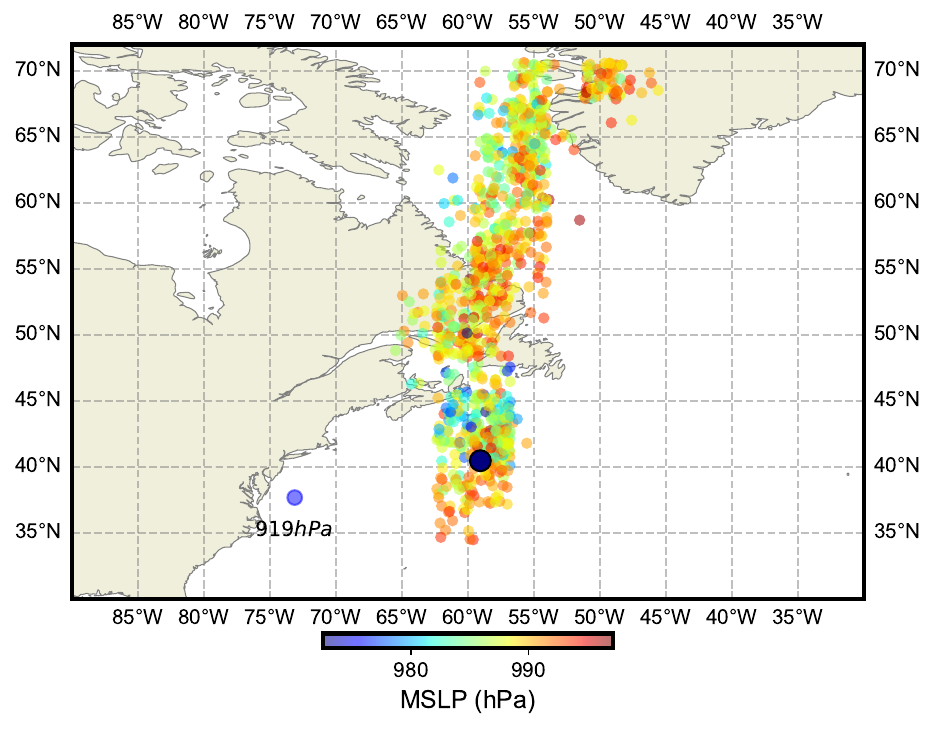}
\end{center}
  \caption{Forecast storm minimum MSLP for a 1000-member ensemble initialized 22/00UTC valid 24/00UTC. The latitude and longitude locations of the storm minimum MSLP are randomly perturbed by uniform random numbers drawn on the interval [$-1.4$,1.4] (the grid scale) so that all points are visible. Colored circled show the minimum MSLP for each ensemble member, with the control solution denoted by a larger circle. The gray swan solution is shown by the light blue dot (919hPa). Random initial conditions for the forecasts are defined in the text.} \label{fig:random_ensemble}
\end{figure}

Although the gray swan solution allows one to test ``what if'' hypotheses about extreme events, essentially testing worst-case scenarios, these solutions do not inform on the probability of such outcomes. An illustration that the Fiona gray swan is unlikely is shown in Fig.~\ref{fig:random_ensemble}, summarizing forecasts at 24/00UTC for a 1000-member ensemble derived from randomly perturbing the ERA5 temperature field at 22/00UTC. The initial perturbations are randomly drawn in spectral space with a mean-squared amplitude that is the same as for the gray swan averaged over the region 40W--60W, 20--60N. Evidently, random perturbations yield weaker storms, with spread mostly in the along-track storm position with less spread across the track. No ensemble member remotely approximates the position or intensity of the gray swan solution, reflecting the fact that the gray swan is an outcome of an efficient targeted search algorithm rather than a random outcome. 

Another measure of the likelihood of the Fiona gray swan solution derives from comparing the forecast 500 hPa temperature field in the upstream trough, which measures its intensity in a way that can be compared directly to observations. Recall that this trough is a primary target for the optimization routine due to its interaction with the hurricane. In the gray swan forecast, the minimum 500hPa temperature in the trough is $-19^{\circ}$C at 23/00UTC and $-22^{\circ}$C at 23/12UTC during the onset of rapid deepening; it increases to $-16^{\circ}$C at 23/18UTC. Radiosonde observations from Wallops Island, Virginia, show a record minimum 500hPa temperature of $-21^{\circ}$C for 23/12UTC (1963--2025). At Sterling, Virginia, the record minimum is $-20^{\circ}$C for 23/12UTC and $-23^{\circ}$C for 21/00UTC (1960--2025). These results suggest that the intensity of the precursor trough is near the limit of the observational record for this time of the year. Combined with an estimate of the probability of a hurricane coincident with such a trough might provide a way to estimate the probability of this gray swan outcome.

\begin{figure}[ht!]
\begin{center}
  \noindent\includegraphics[width=1.0\textwidth]{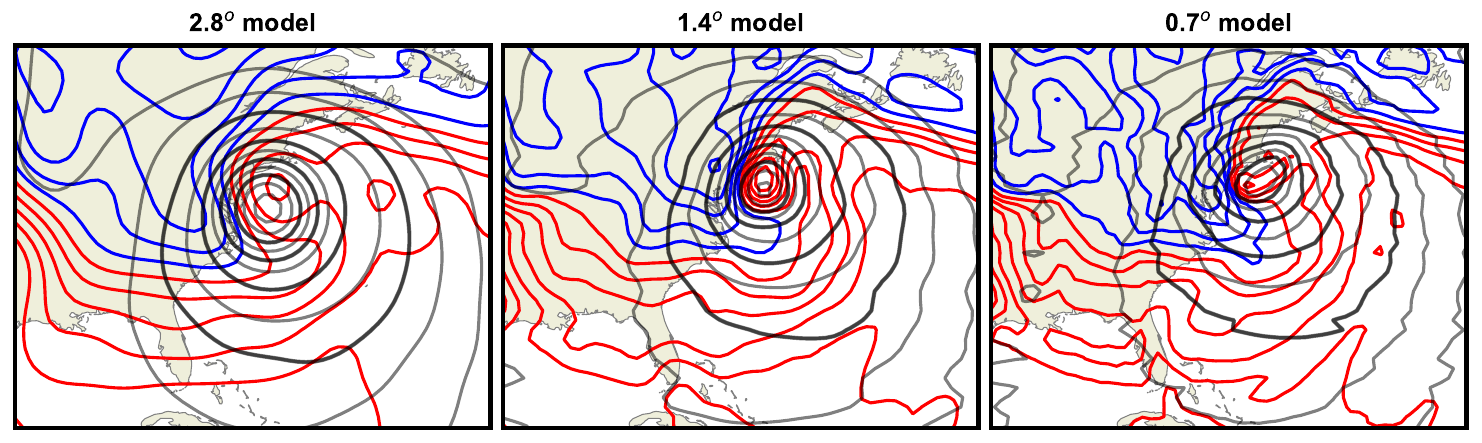}
\end{center}
  \caption{Forecasts initialized 22/00UTC with the gray swan initial condition, valid 24/00UTC, as a function of NGCM model resolution. MSLP contours are shown every 8 hPa, with darker contours for 944, 960, 976, and 992 hPa. 1000hPa isotherms are shown every 2K with red for values at and above 292K, and blue otherwise. Storm minimum PMSL for each model: 919 hPa for 2.8$^{\circ}$; 927 hPa for 1.4$^{\circ}$; 929 hPa for 0.7$^{\circ}$. Storm minimum MSLP locations are within one grid cell of the 2.8$^{\circ}$ model.} \label{fig:model_resolution}
\end{figure}

\section{Sensitivity tests \label{sec:sensitivity}}

\begin{figure}[ht!]
\begin{center}
  \noindent\includegraphics[width=1.0\textwidth]{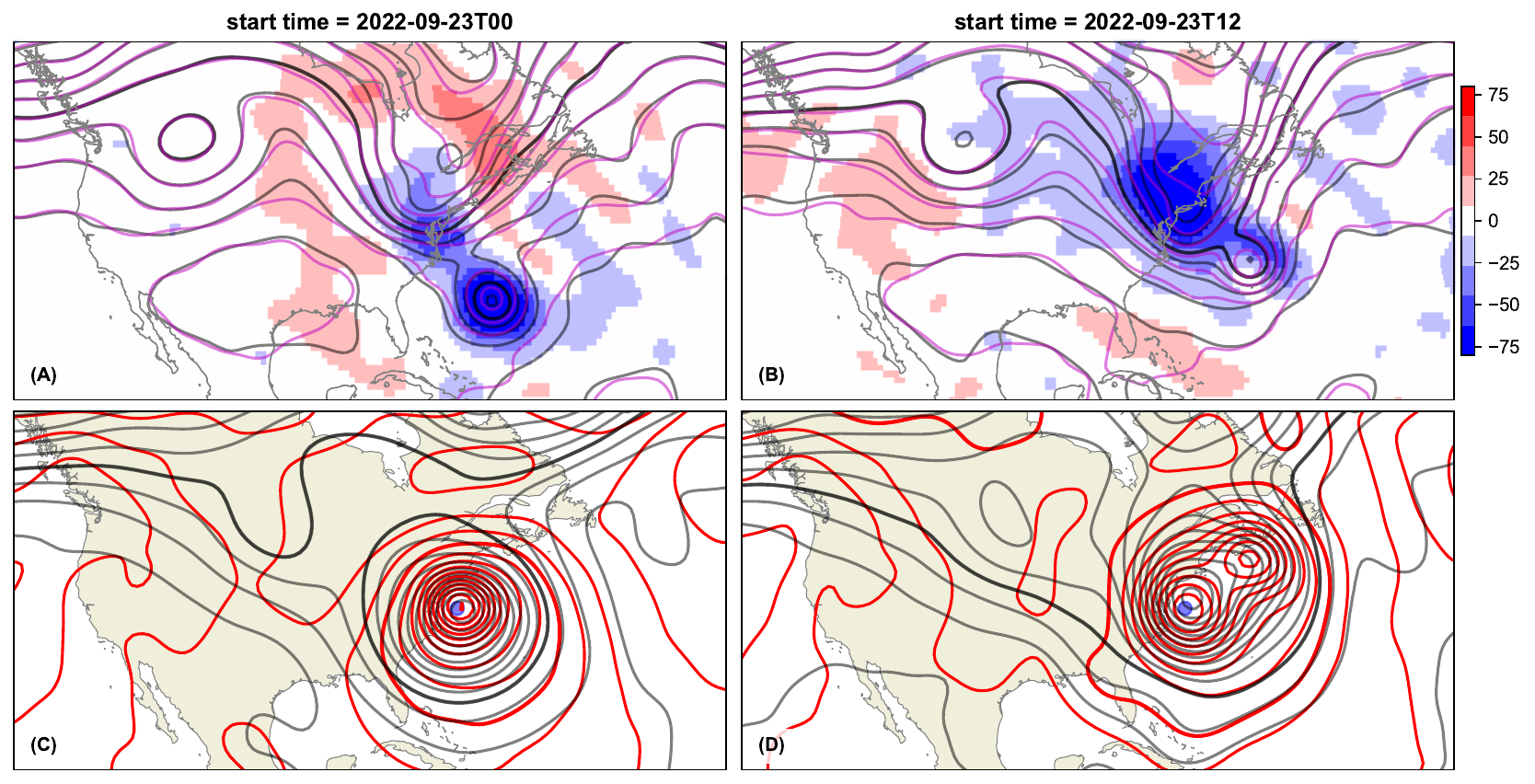}
\end{center}
  \caption{Gray Swan solutions for Fiona starting at (A), (C) 23/00UTC and (B, (D) 23/12UTC and optimized for a Sandy-like outcome 24/00UTC. (A), (B) show the 500 hPa geopotential height for the gray swan (gray lines) and control (magenta lines) every 60m with the 5640m contour bold; gray swan minus the control in colorfill. (C), (D) the gray swan solutions at 24/00UTC showing MSLP in red contours every 8 hPa, with the 992 hPa contour bold; 500 hPa geopotential height is shown in gray contours every 60m with the 5640m contour bold.} \label{fig:optimal_start_time}
\end{figure}

The gray swan solution is found using a coarse 2.8$^{\circ}$ model, and one natural concern is the sensitivity of the solution to model resolution. To test this sensitivity, Fig.~\ref{fig:model_resolution} shows the solution for the 1.4$^{\circ}$ and 0.7$^{\circ}$ degree versions of NGCM initialized with the gray swan initial condition by downscaling using spectral interpolation (zero padding at unresolved wavenumbers). Results show that the storm is slightly weaker in the high resolution simulations (929 hPa at  0.7$^{\circ}$ compared to 919 hPa at  2.8$^{\circ}$ resolution), but still stronger than hurricane Sandy, and the storm location in the higher-resolution solutions is within one grid cell of the coarse model. Furthermore, the structure is similar across model resolutions, but with larger gradients at higher resolution. This is particularly notable in the warm core seclusion and bent-back warm front, both hallmarks of extreme extratropical cyclones in general \cite{shapiro_keyser}, and Sandy (2012) in particular \citep{galarneau2013intensification}. All solutions show a warm-core cyclone, but the local maximum at 2.8$^{\circ}$ resolution is limited to one 2K contour, displaced from the MSLP minimum. The strongest warm core is evident in the 1.4$^{\circ}$ solution, 6K warmer than the surroundings, at least hinting at a hurricane-like inner core similar to Sandy. At 0.7$^{\circ}$, the elliptical structure of the MSLP field near the low center suggests the possibility of a double low at even higher resolution. A bent-back warm front is evident in all solutions, with an increase in the magnitude of the horizontal temperature gradient with resolution. This lack of sensitivity to model resolution is encouraging because these versions of NGCM are different models in the sense that, in addition to the resolution differences, they have separately trained learned components that are aware of the resolution differences. Nevertheless, it is important to test the robustness of the gray-swan solutions in higher-resolution traditional models, which is beyond the scope of this paper and a subject of future research. In particular, it will be very interesting to explore how the solution changes when the inner core of hurricane is resolved.

One design choice in computing the gray swan solution is the length of time between the initial and optimization times. For the standard configuration, we optimize over the 48h period from 22/00UTC to 24/00UTC, and now explore the sensitivity of the results to changing the starting time. We find little sensitivity for changes within $\sim$24h of 22/00, and an increase in sensitivity for shorter time intervals (Fig.~\ref{fig:optimal_start_time}). While the solution for 23/00UTC  (Fig.~\ref{fig:optimal_start_time}A,C) is very similar to that for 24/00UTC, the solution for 23/12 yields two separate storms, one at the optimization location and the other on a track similar to the control  (Fig.~\ref{fig:optimal_start_time}B,D). The shorter time period does not allow enough time to significantly affect the track of Fiona, which shifts the emphasis of the initial perturbations to changing the intensity of the 500~hPa upstream trough (cf. Fig.~\ref{fig:optimal_start_time}A,B), and the formation of an extratropical surface cyclone along the coast, independent of the hurricane.

\section{Gray swans for other storms \label{sec:other_storms}}
\begin{figure}[ht!]
\begin{center}
  \noindent\includegraphics[width=.95\textwidth]{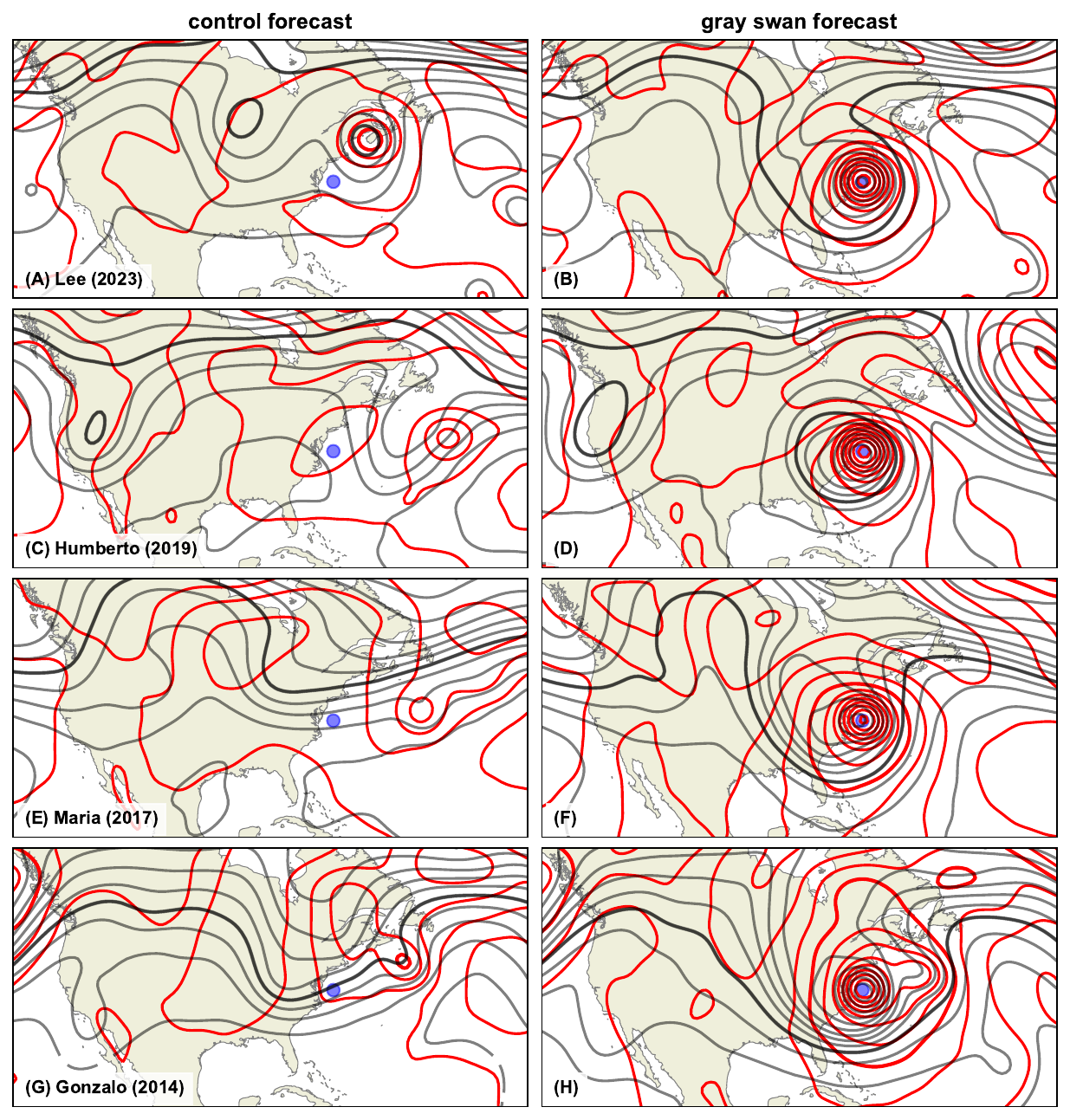}
\end{center}
  \caption{Control (left panels) and gray swan (right panels) forecasts for storms listed in Table~\ref{table:swans}. optimals for other storms. MSLP is shown in red contours every 8 hPa, with the 992 hPa contour bold, and 500 hPa geopotential height is shown in gray contours every 60m with the 5640m contour bold. Starting and ending times for the forecasts are given in Table~\ref{table:swans}. The blue dot shows the optimization location at the end time for each storm.} \label{fig:optimal_other_storms}
\end{figure}

\begin{figure}[ht!]
\begin{center}
  \noindent\includegraphics[width=.95\textwidth]{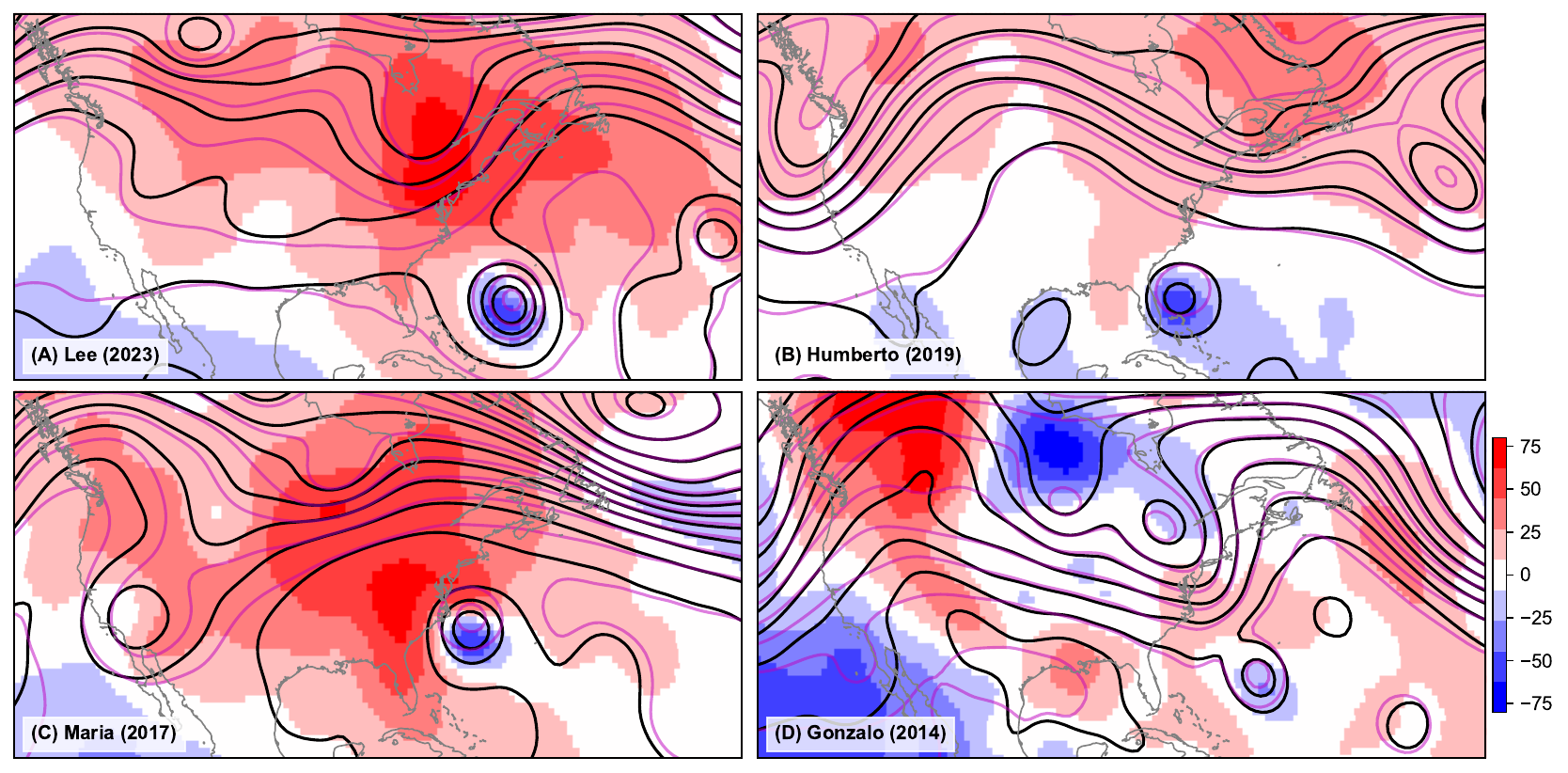}
\end{center}
\caption{Initial conditions for the forecasts shown in Figure~\ref{fig:optimal_other_storms}. The 500 hPa geopotential height is shown every 60m with the 5640m contour bold for the gray swan in gray lines and the control in magenta lines. The difference field (gray swan minus the control) is shown in colorfill.}
\label{fig:optimal_ics_other_storms}
\end{figure}

To assess whether the gray swan outcome is peculiar to Fiona (2022), we apply the standard configuration (1000 epochs with all hyperparameters fixed at those for Fiona) to four other hurricanes (Table \ref{table:swans}; Fig.~\ref{fig:optimal_other_storms}). The storms are subjectively chosen to have tracks similar to Sandy and Fiona, moving northward over the western Atlantic during September and October during recent years. Results show that all storms yield Sandy-like gray swans, with minimum MSLP less than Sandy's 940~hPa. The change in the end-time minimum MSLP between the gray swan and the control solution ranges from $-48$~hPa for Humberto to $-64$~hPa for Gonzalo. The initial intensity difference for the gray swans is notably smaller for all storms compared to Fiona, ranging from 0--4 hPa compared to 8~hPa for Fiona (Table \ref{table:swans}). In contrast, changes to the initial extratropical synoptic features at 500 hPa are similar or larger when compared to Fiona (Fig.~\ref{fig:optimal_ics_other_storms}). It appears that, for the storms considered, the main factor controlling the intensity of the gray swan is the structure and intensity of the extratropical flow that subsequently interacts with the hurricane. This could be due to the coarse model resolution used for these calculations, which doesn't resolve hurricane structure or intensity. Alternatively, the fact that Sandy-like storms involve interactions with troughs in the jet stream could mostly require small changes in trough phase and amplitude to facilitate the merger of these systems.

%

\begin{table}[htbp]
    \caption{Gray swan summary. Times are shown in year-month-day format, and storm minimum pressure in hPa.}
    \centering
    \begin{tabular}{lllllll}
        \toprule
        & \multicolumn{2}{c}{Optimization time window} & \multicolumn{2}{c}{Control MSLP} & \multicolumn{2}{c}{Swan MSLP} \\
        \cmidrule(r){2-3} \cmidrule(r){4-5} \cmidrule(r){6-7}
        Storm (year) & Start & End & Start & End & Start & End \\
        \midrule
        Fiona (2022) & 2022-09-22 & 2022-09-24 & 987 & {\bf 976} & 979 & {\bf 919} \\
        Lee (2023) & 2023-09-14 & 2023-09-17 & 982 &  {\bf 985} & 981 &  {\bf 929} \\
      Humberto (2019) & 2019-09-16 & 2019-09-20 & 1000 &  {\bf 984} & 1000 &  {\bf 936} \\
        Maria (2017) & 2017-09-27 & 2017-09-30 & 987 &  {\bf 989} & 987 &  {\bf 930} \\
        Gonzalo (2014) & 2014-10-17 & 2014-10-19 & 965 &  {\bf 986} & 961 &  {\bf 922} \\
      
        \bottomrule
    \end{tabular}
    \label{table:swans}
\end{table}
   
The gray swan solutions at the end time are distinctly different (Fig.~\ref{fig:optimal_other_storms}) across the sample of storms. By design, all cases yield a storm on the coast, similar to Fiona, but the downstream and upstream ridges in the 500hPa height field in particular differ in amplitude and scale. The control solutions have a wide range of phase relationships between the tropical cyclone and an upstream trough, which appears to be the key feature responsible for Fiona's gray swan. This is also evident in the initial conditions (Fig.~\ref{fig:optimal_ics_other_storms}), which show a wide range of extratropical synoptic variability. The gray swan initial conditions are all associated with modestly stronger initial storms. with a shift in position. Maria and Lee in particular show shifts in storm position (Fig.~\ref{fig:optimal_ics_other_storms}A,C), which the optimization process has identified as important for subsequent phasing with an extratropical troughs and a transition to extratropical cyclone development. 

\section{Conclusions \label{sec:conclusions}}

Gray swans are extreme events that are plausible, but not previously observed. We use this concept to explore the possibility of finding initial conditions close to a reference, such as a reanalysis product, with forecast outcomes that optimize a performance metric. This approach falls within the general topic of reachability in dynamical systems theory. We use a version of the optimization technique described in \citet{whittaker2026} to manufacture extreme cyclones like hurricane Sandy (2012) from ``ordinary'' hurricanes. Hurricane Sandy is unique in the observational record for a westward turn that changed its course from a typical offshore trajectory to one producing landfall on the East Coast of North America.  Our loss function has the objective of lowering the surface pressure near the location where Sandy made landfall, and the optimization process starts with an initial condition drawn from a reanalysis dataset for five subjectively chosen storms that remained offshore in reality. Optimization of the loss is performed over 2--4 days depending on the storm, and the loss function also penalizes the size of the perturbations to the initial condition, effectively limiting the range of reachable states.

Our main finding is that, for a sample of five storms, the optimization process yields gray swan cyclones in the target location of Sandy's landfall with intensity greater than Sandy as measured by minimum mean-sea-level pressure. Sensitivity of the results to various design choices was explored for the gray swans of hurricane Fiona (2022). An important sensitivity involves the initial-condition penalty function, which yields the control solution in the limit of infinite penalty, and a deeper storm (minimum pressure less than 900 hPa) in the limit of zero penalty. The length of time for the optimization also affects the range of outcomes by limiting the reachable states. For Fiona, reducing the window to 12h yields two storms, one extratropical and one tropical, rather than a single hybrid storm resulting from the merger of a hurricane with an extratropical cyclone. Another test explored the contribution of initial-condition perturbations to the final state revealed that most of the gray swan Sandy-like outcome derived from changes to the extratropical circulation, such as the location and intensity of a trough in the jet stream. These changes appear to affect the timing and details of the interaction with the hurricane leading to the Sandy-like gray swan.

These results appear to suggest that the atmosphere's attractor is dense with Sandy-like outcomes ``near'' ordinary storms. The probability of such outcomes is unknown, although an ensemble experiment for Fiona starting from 1000 random initial-condition perturbations yields no forecasts remotely similar to the gray swan. This reflects the highly specific changes to the initial conditions, which apparently would require much larger random ensembles to discover with a stochastic, non-targeted, approach. We emphasize the limitations of these proof-of-concept solutions that are derived using a model with coarse spatial resolution and a learned component for unresolved physics, rather than a high-resolution traditional model with process-specific parameterization schemes. In particular, the structure of the hurricane is unresolved in our simulations, and the apparent importance of the extratropical perturbations could derive mainly from that limitation. Simulations using the gray swan initial conditions in a high-resolution traditional model are important tests of these results. Moreover, similar experiments using different differentiable models is another important topic for future research, which would test the robustness of the solutions presented here. Another fruitful direction involves testing the ability of purely data-driven models to learn from synthetic datasets having controllable amounts of gray-swan training data generated by the method described here.

\section{Acknowledgements\label{sec:ack}}
Radiosonde data were obtained from {\tt www.spc.noaa.gov/exper/soundingclimov2}. We acknowledge support from NSF awards 2501400 and 2530556, and Heising-Simons Foundation award 2023-4715 made to the University of Washington. 

\section{Author Contributions}
GJH conceived of the study, performed all calculations, and wrote the first draft of the paper. AA gathered data on tropical cyclones for candidate selection and helped source background literature. Both authors contributed to the final draft. 

%

\bibliographystyle{unsrtnat}
\bibliography{references}  






\end{document}